\documentclass[preprint2]{aastex6}
\usepackage[fleqn]{amsmath}
\usepackage{amssymb}

\usepackage[normalem]{ulem}

\slugcomment{doi: \url{https://doi.org/10.3847/1538-3881/aaaaed}}

\shortauthors{J. Pe\~na et al.}

\begin{document}


\title{Asteroids in the High cadence Transient Survey}


\author{J. Pe\~na\altaffilmark{1, 2, 3},
  C. Fuentes\altaffilmark{1,3},
  F. F\"orster\altaffilmark{2, 3},
  J.C. Maureira\altaffilmark{2},
  J. San Mart\'in\altaffilmark{2},
  J. Litt\'in\altaffilmark{4},
  P. Huijse\altaffilmark{3,5},
  G. Cabrera-Vives\altaffilmark{6,3,2,7},
  P.A. Est\'evez\altaffilmark{5,3},
  L. Galbany\altaffilmark{8},
  S. Gonz\'alez-Gait\'an\altaffilmark{2, 3, 9},
  J. Mart\'inez\altaffilmark{1, 2, 3},
  Th. de Jaeger\altaffilmark{10, 1, 3},
  M. Hamuy\altaffilmark{1, 3}
}
\email{jpena@das.uchile.cl}


\altaffiltext{1}{Departamento de Astronom\'ia, Universidad de Chile, Camino del Observatorio 1515, Las Condes, Santiago, Chile.}
\altaffiltext{2}{Center for Mathematical Modelling, Beaucheff 851, 7th floor, Santiago, Chile.}
\altaffiltext{3}{Millennium Institute of Astrophysics, Chile.}
\altaffiltext{4}{Departamento de Matem\'aticas, Universidad Cat\'olica del Norte, Angamos 0610,
Antofagasta, Chile.}
\altaffiltext{5}{Electrical Engineering Department, University of Chile.}
\altaffiltext{6}{Department of Computer Science, Universidad de Concepci\'on.}
\altaffiltext{7}{AURA Observatory in Chile.}
\altaffiltext{8}{PITT PACC, Department of Physics and Astronomy, University of Pittsburgh, Pittsburgh, PA 15260, USA.}
\altaffiltext{9}{CENTRA, Instituto Superior T\'ecnico, Universidade de Lisboa, Portugal.}
\altaffiltext{10}{Department of Astronomy, University of California, Berkeley, CA 94720-3411, USA.}


\begin{abstract}
We report on the serendipitous observations of Solar System objects imaged during the High cadence Transient Survey (HiTS) 2014 observation campaign. Data from this high cadence, wide field survey was originally analyzed for finding variable static sources using Machine Learning to select the most-likely candidates. In this work we search for moving transients consistent with Solar System objects and derive their orbital parameters.

We use a simple, custom detection algorithm to link trajectories and assume Keplerian motion to derive the asteroid's orbital parameters. We use known asteroids from the Minor Planet Center (MPC) database to assess the detection efficiency of the survey and our search algorithm.
Trajectories have an average of nine detections spread over 2 days, and our fit yields typical errors of $\sigma_a\sim 0.07 ~{\rm AU}$, $\sigma_{\rm e} \sim 0.07 $ and $\sigma_i\sim 0.^{\circ}5~ {\rm deg}$ in semi-major axis, eccentricity, and inclination respectively for known asteroids in our sample.  We extract 7,700 orbits from our trajectories, identifying 19 near Earth objects, 6,687 asteroids, 14 Centaurs, and 15 trans-Neptunian objects. 
This highlights the complementarity of supernova wide field surveys for Solar System research and the significance of machine learning to clean data of false detections. It is a good example of the data--driven science that LSST will deliver.

\end{abstract}



\section{Introduction}

We are entering the age of wide-field surveys, where massive amounts of data are collected to gather information on as many sources as possible while monitoring how the sky changes over time. In this new era, data can be refurbished to be useful for a science case different from the one it was originally designed for. This is particularly true for ``time domain'' astronomy, where multiple observations of a region are obtained in hopes of finding differences between data taken in different epochs. For example this has been used in the past to search for moving objects in the Hubble Space Telescope's archive: \cite{Fuentes.2010, Fuentes.2011}, where most of the objects were discovered in data obtained for SNe characterization. Other surveys not mainly focused in Solar System science, such as SDSS \citep{York.2000}, WISE \citep{Wright.2010} or PS-1 \citep{Chambers.2016} had also been used for detection and study of Solar System's Minor Planets (see Figure \ref{table:survey} for more examples).

Supernovae (SNe), Solar System (SS) objects, or transiting Exoplanets all need multi-epoch observations to be discovered. The timescale over which the observations must be carried out depends, however, on the particular object of study. Hence, while SN surveys usually observe a field over a couple of weeks, SS objects can be detected in a single night and confirmed a few days later, and planets may need weeks of dense monitoring to detect the ephemeral transit. A survey's success is determined in part by an appropriate choice of cadence, which also determines its value for surveying other phenomena. This is one of the aspects that has to be resolved for the largest survey to come: The Large Synoptic Survey Telescope (LSST) project. The LSST will marry different communities by devoting its operation to imaging the sky efficiently and consistently. In order to do so, a great deal of effort is being spent in selecting the best observing strategy to fulfill the science needs of all the groups involved.

In this paper we present our search for Solar System objects in the High cadence Transient Survey (HiTS), a high cadence survey designed to find young Supernovae using the Dark Energy Camera (DECam) at the 4m Blanco Telescope on Cerro Tololo Observatory (\citealt{2016ApJ...832..155F}, hereafter F16). DECam has been hailed as a precursor of LSST for its large field of view, fast readout, and large 520 Mpix CCD camera \citep{DECAM}.

In this work we look for SS objects in HiTS 2014 data, which contains several observations of the same area of the sky during consecutive nights. We test the expected discovery efficiency for the known population of asteroids, while extending the sensitivity to smaller objects by at least one magnitude. We show that high cadence Supernovae surveys in general are well suited to search and characterize the orbits of SS objects. Our results for this sample of small asteroids are consistent with those of asteroids already studied.

The HiTS data is described in section 2. In section 3 we show the linking algorithm used to find asteroids and other SS objects. Our results and their comparisons with known objects is presented in section 4. We discuss the results and provide ideas for future work in Section 5.

\section{Data}

\subsection{HiTS observations}

HiTS was run in three different campaigns in the 2013A, 2014A and 2015A semesters. The 2013A campaign consisted of four consecutive nights of $u$--band observations of 120 ${\rm deg}^2$ (40 DECam pointings) with a cadence of about two hours (four observations per night). The 2014A campaign consisted of five consecutive nights of $g$--band observations towards 120 ${\rm deg}^2$ (40 DECam pointings) with a cadence of about two hours (four observations per night). The 2015A campaign consisted of 6 consecutive nights of mainly $g$--band observations and a few observations in the $r$ and $i$ bands, towards 150 ${\rm deg}^2$ (50 DECam pointings) with a cadence of about 1.6 hours (five observations per night), followed by three non--consecutive half nights 2, 5 and 20 nights after the end of the main run. The location of the HiTS fields is shown in Figure 4 of F16. In this work we only analyze data from the 2014A campaign, which is the deepest of the previous three campaigns.

\subsection{Data Processing} \label{sec:dataproc}

The HiTS survey runs a custom--made pipeline in real--time to detect fast transients. Given its cadence, with several observations per night, and the location of its fields, close to the Sun's opposition in 2014, it was ideally suited for the detection of asteroids as well, which we analyze in this publication. 

The raw data was pre--processed with a local copy of the DECam Community Pipeline (DCP), which includes electronic bias calibrations, crosstalk corrections, saturation masking, bad pixel masking and interpolation, bias calibration, linearity correction, flat field gain calibration, fringe pattern subtraction, bleed trail and edge bleed masking, and interpolation (F16).

  Pre--processed data was template subtracted with the HiTS pipeline described in F16 and \cite{Cabrera.2017}. Images are registered using a Lanczos 2 kernel and convolved using a variable pixel size kernel. After differencing, variable candidates are detected using the optimal photometry method (\citealt{Naylor.1998}), and classified as real or bogus using a random forest classifier (RF), which retrieves a probability for each candidate of being real. The RF was previously trained using data from the 2013A campaign. The classification of sources requires the computation of the candidates' features, which absorbs most of the computational cost. Calculating the 56 features in a single 2.2 GHz processor takes $\sim$11 seconds for $\sim$5,000 candidates.

Selecting only those candidates with probability higher than .5, the data is reduced by $\sim$80\%, reducing our original $1.8$ million candidates in HiTS to $\sim$360,000.

Since convolution forces the same calibration for the template and science images, our photometry is based on the absolute calibration of the template images. We estimate our photometric precision in the $g$ band to be better than 0.02 mag. for objects brighter than 21 mag. and to increase up to 0.2 mag. at 23 mag. (Mart\'inez et al. \emph{submitted}).

\subsection{Survey Efficiency}
We use the known asteroid population from the Minor Planet Center \footnote{Information provided via web page in \url{http://www.minorplanetcenter.net/cgi-bin/checkmp.cgi}} (MPC) as an unbiased sample to test the asteroid detection efficiency of the HiTS survey. By checking the number of times a known asteroid is detected as a variable source we get an accurate assessment of the maximum number of asteroids our linking algorithm can identify as a moving object.

\begin{figure}
   \centering
   \includegraphics[width=\hsize]{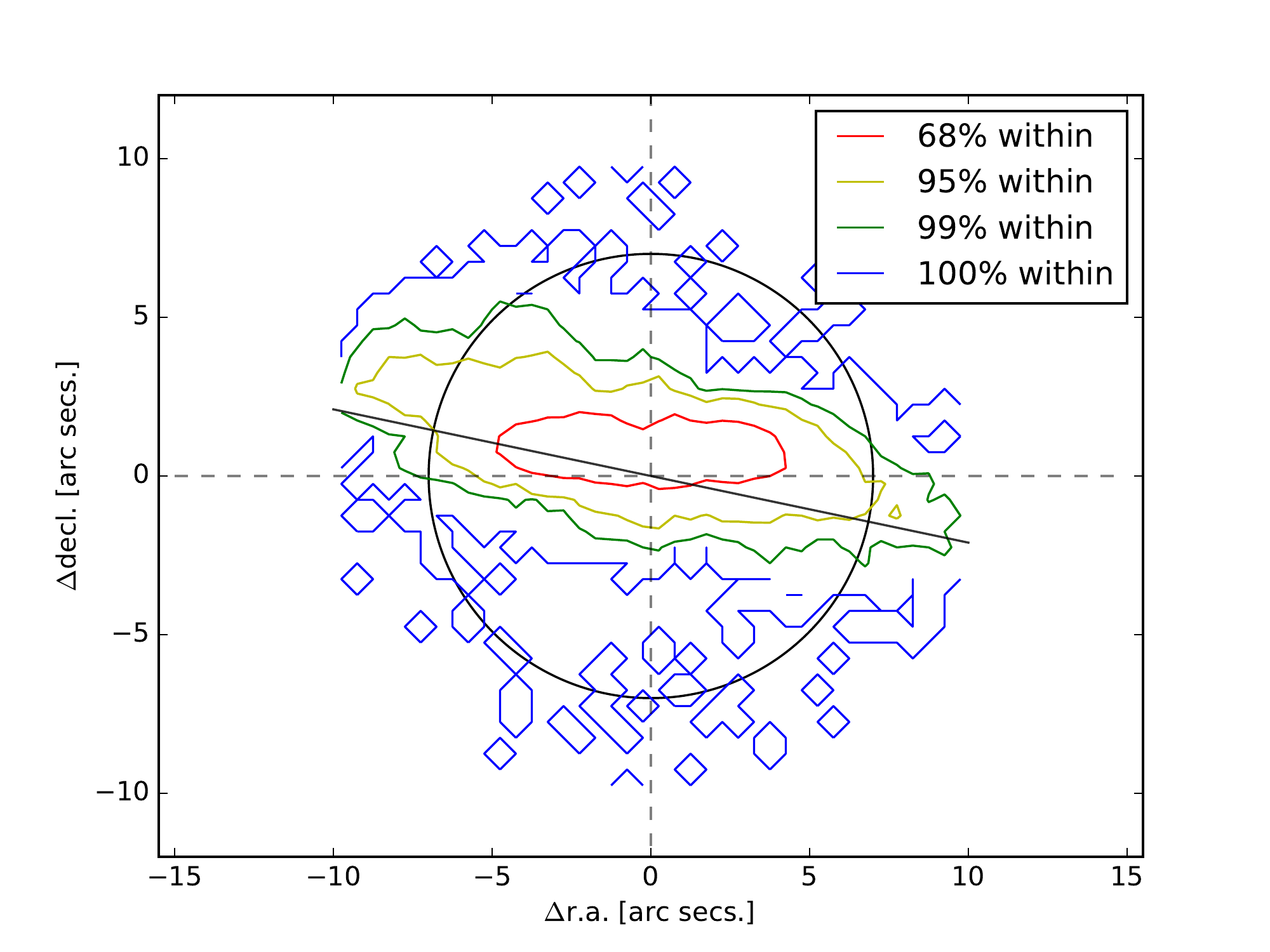}
   \caption{Contour plot for the difference between coordinates of MPC and HiTS data. Each contour surrounds a percentage of the matched data. Straight black line has the same slope as the ecliptic. Black circle has a 7 arc-seconds radius.}
   \label{fig:distcontour}%
\end{figure}

We match HiTS's detections with known asteroids for every epoch, measuring their projected separation in the sky as shown in Figure \ref{fig:distcontour}. We note that most asteroids are identified with a detection lying within $\sim$7 arc-seconds, while the position errors are larger along the ecliptic. Based on this plot we consider a $7''$ tolerance in projected distance to match detections in HiTS with MPC's positions of known asteroids. The typical sky density of our variable source list is $\sim 120\rm{deg}^{-2}$, while the density of the MPC list of asteroids neighboring our pointing is $40\rm{deg}^{-2}$, which brings our chances of getting one spurious match per square degree down to 6\%. In consequence, this is the same probability of including one false detection on an otherwise fully correct track, per square degree (see section 3 for the definitions of \emph{detection} and \emph{track}). Because there can be several asteroids per square degree during the five days of our survey is not unusual to have trajectories that link real detections of asteroids with a few false detections. We call this \emph{track confusion}. Beside this, from simulations of one field (using the same characteristics of one of ours, but with randomly simulated detections) we find that the probability of linking an entire track with unrelated detections (a false track) is less than 2\%, resulting in tracks that almost certainly would be rejected when doing the keplerian fit (see section 4.2).

\begin{figure}
   \centering
   \includegraphics[width=\hsize]{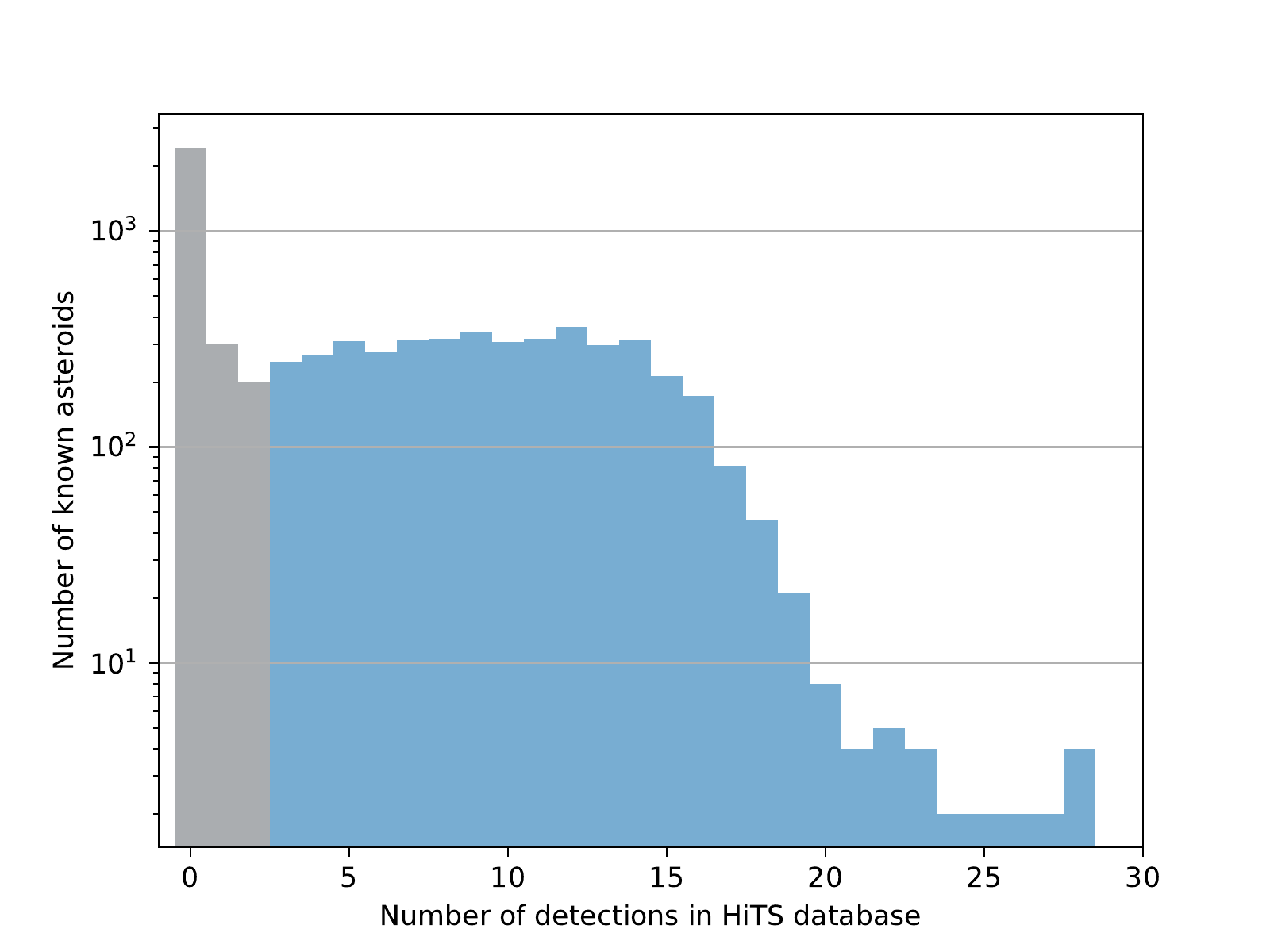}
   \caption{Histogram of number of times an asteroid was imaged and analyzed as a variable candidate source by the HiTS survey. All known asteroids (from MPC) that were within 1.25 degrees of any DECam pointing in the HiTS survey during 2014 were considered in this analysis. Asteroids with less than three detections are shown in gray, those with three and more are colored blue. }
   \label{fig:detectedhist}%
\end{figure}

We considered $\sim 7,700$ different known asteroids near HiTS pointings. In Figure \ref{fig:detectedhist} we show the number of times one of these known asteroids is matched with a detection in the HiTS variable source list (projected distance under $7''$). We divide the sample in those that are matched under three times (gray) and those that have enough detections to be linked (blue). The detection efficiency does not improve significantly when using the entire $1.8$ million candidates from HiTS (without the machine learning selection), losing 5\% of the known asteroids' detections, which translates into a 2\% drop in the number of known asteroids that we are able to recover and identify.

\section{Analysis}

The data used for this work consists of catalogs of variable object candidates obtained via image subtraction as explained in Section~\ref{sec:dataproc}. The catalogs were filtered using a random forest classifier based on candidate image stamps (see F16). Candidates with a probability larger than 0.5 of being real according to the classifier are called \emph{detections} in this work.
We link detections into trajectories using two different algorithms. The first one works as follows: we first find linear segments of at least 5 detections by looking for clusters of relative velocities between detections. Then we join linear segments to form more complete curved trajectories. These trajectories are also processed to eliminate outliers or to add detections not found in the previous steps. Finally, we remove trajectories with a high acceleration and we force trajectories to be disjoint sets. These trajectories are actually candidates for real trajectories, so we just call them \emph{tracks}. The execution time of this algorithm scales as ${\cal O}(n^{2})$. This means that if we wouldn't have used Machine Learning to reject $\sim 80\%$ of the original data, it would have taken 25 times longer to find our tracks.  When comparing to the list of known asteroids this algorithm yields about a 30\% efficiency.
The second linking algorithm is similar to the one used in \cite{Trilling.2017} and \cite{Valdes.2015}, with an execution time that scales as ${\cal O}(n \log n)$. As this algorithm runs much faster than the first one, we considered three-detection tracks, improving the linking efficiency as shown in the next section.
When considering tracks of at least 5 detection the efficiency is identical for both algorithms. Since there are at most 4 detections for the same object in a single night, requiring at least 5 detections leaves many single-night tracks out. Looking for single-night tracks and then joining them yields better results (see Figure \ref{fig:eff}).

\section{Results}

\subsection{Magnitude Distribution}
\begin{figure}
   \centering
   \includegraphics[width=\hsize]{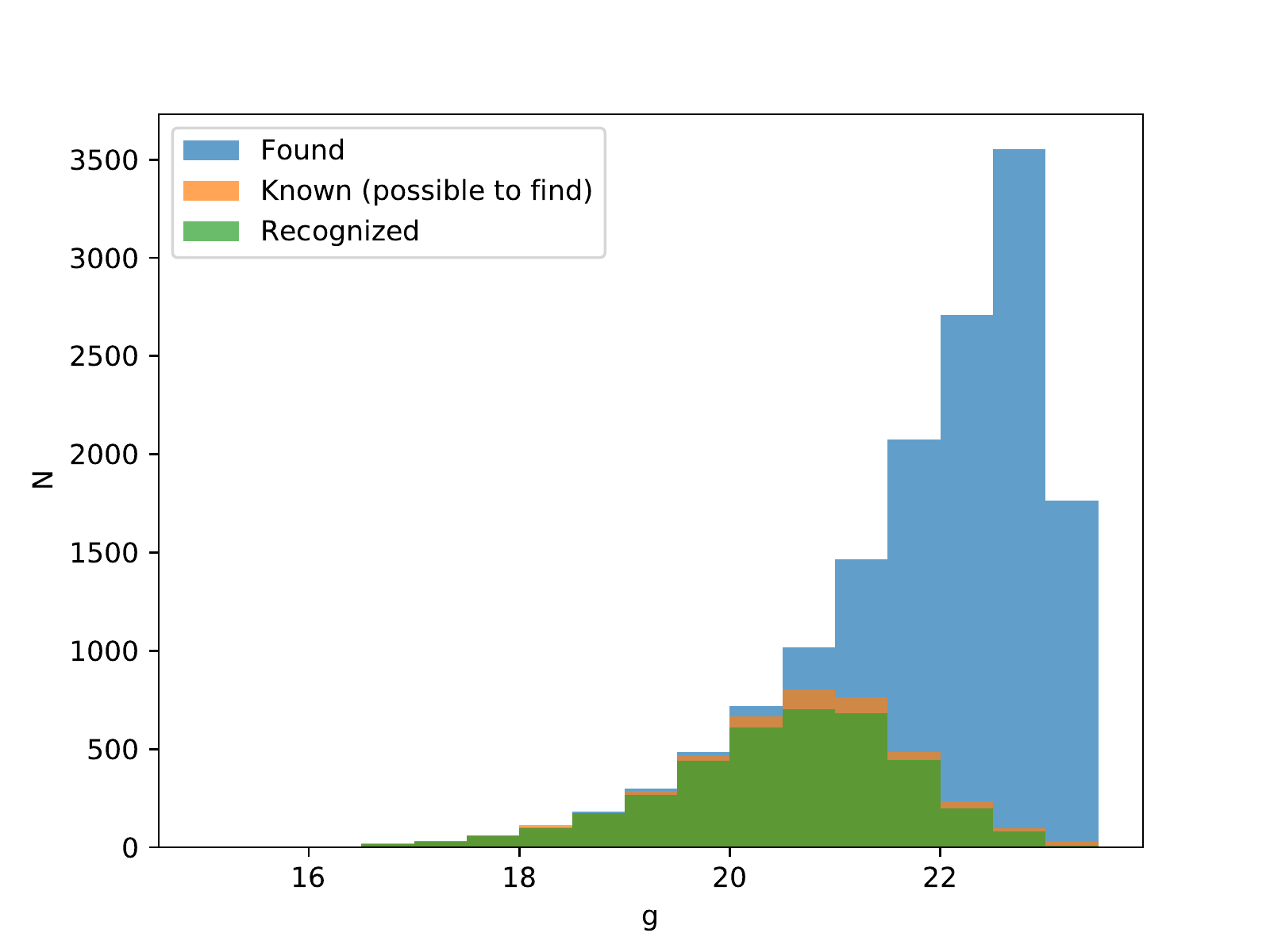}
   \caption{Histograms of number of asteroids per magnitude (g band). In orange, known asteroids (from MPC) that our linking process can find (have three matches with our data). In blue, tracks found in the HiTS data. In green, those tracks that were recognized as known objects. The drop in the blue histogram appears consistent with Figure~7 in F16 once the effect of image subtraction is taken into account, which results in a loss of $\sim$0.4 mag.}
   \label{fig:maghist}%
\end{figure}

The second algorithm presented in the previous section yielded a total of 14,507 tracks. In Figure~\ref{fig:maghist} we show all these tracks as a blue magnitude histogram, all known asteroids that are present in at least three detections in our data as an orange histogram, and all known asteroids that were also detected as tracks as a green histogram. 
To identify a track as a known asteroid we require at least three detections in the track closer than seven arc-seconds to the same MPC object, which results in 3,812 tracks recognized as known asteroids.
Thus, all objects in the green distribution are also in the blue and orange distributions of Figure~\ref{fig:maghist}.

\begin{figure}
   \centering
   \includegraphics[width=\hsize]{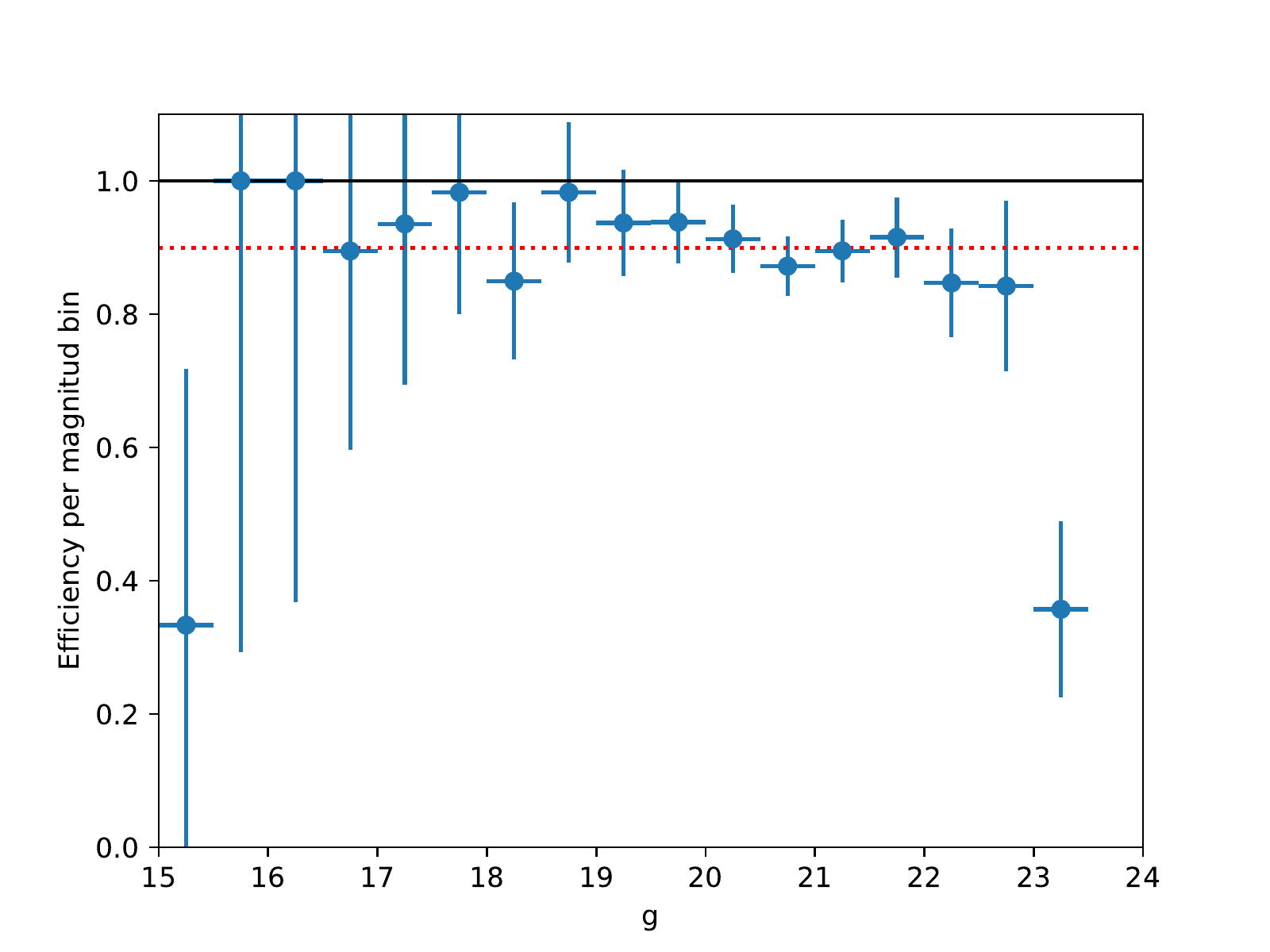}
   \caption{Efficiency of recognized objects over the known objects per magnitude bin (see green and orange histograms in Figure \ref{fig:maghist}). Errors are propagated using Poisson errors for each measurement. The dotted red line represents the total efficiency (total number of recognized tracks over the total number of known objects), equal to 0.9.}
\label{fig:eff}%
\end{figure}

With this information we can calculate the efficiency per magnitude bin of our track finding algorithm, as the ratio between the number of recognized tracks and the number of known objects found with at least three detections in our data. This efficiency is shown in Figure \ref{fig:eff}, where the errors in efficiency are computed propagating Poisson errors.
The total efficiency of our track finding algorithm, defined as the ratio between the total number of recognized tracks and the total number of known asteroids with at least three detections, was found to be 0.9. The total efficiency is shown with a red dotted line in Figure~\ref{fig:eff}. Although our sample population (known objects) is in average brighter than our detections, the apparent drop in detection efficiency in Figure \ref{fig:eff} slightly beyond $g=23$ is consistent with the drop in number of recognized tracks in blue (Figure \ref{fig:maghist}).

\subsection{Orbital Fitting}
A Keplerian orbit was fitted to each track. We used a modified version of the code by \cite{Bernstein.2000}, better suited for asteroids,
to fit all the positions from each object into a sky trajectory. In order to reduce the number of spurious detections we fitted tracks which had 6 or more detections in total and with at least 2 detections in each night where they were found. All trajectories with unbound solutions or with maximum residuals at any detected time larger than 2 arc-seconds were also rejected.

\begin{figure}
   \centering
   \includegraphics[width=\hsize]{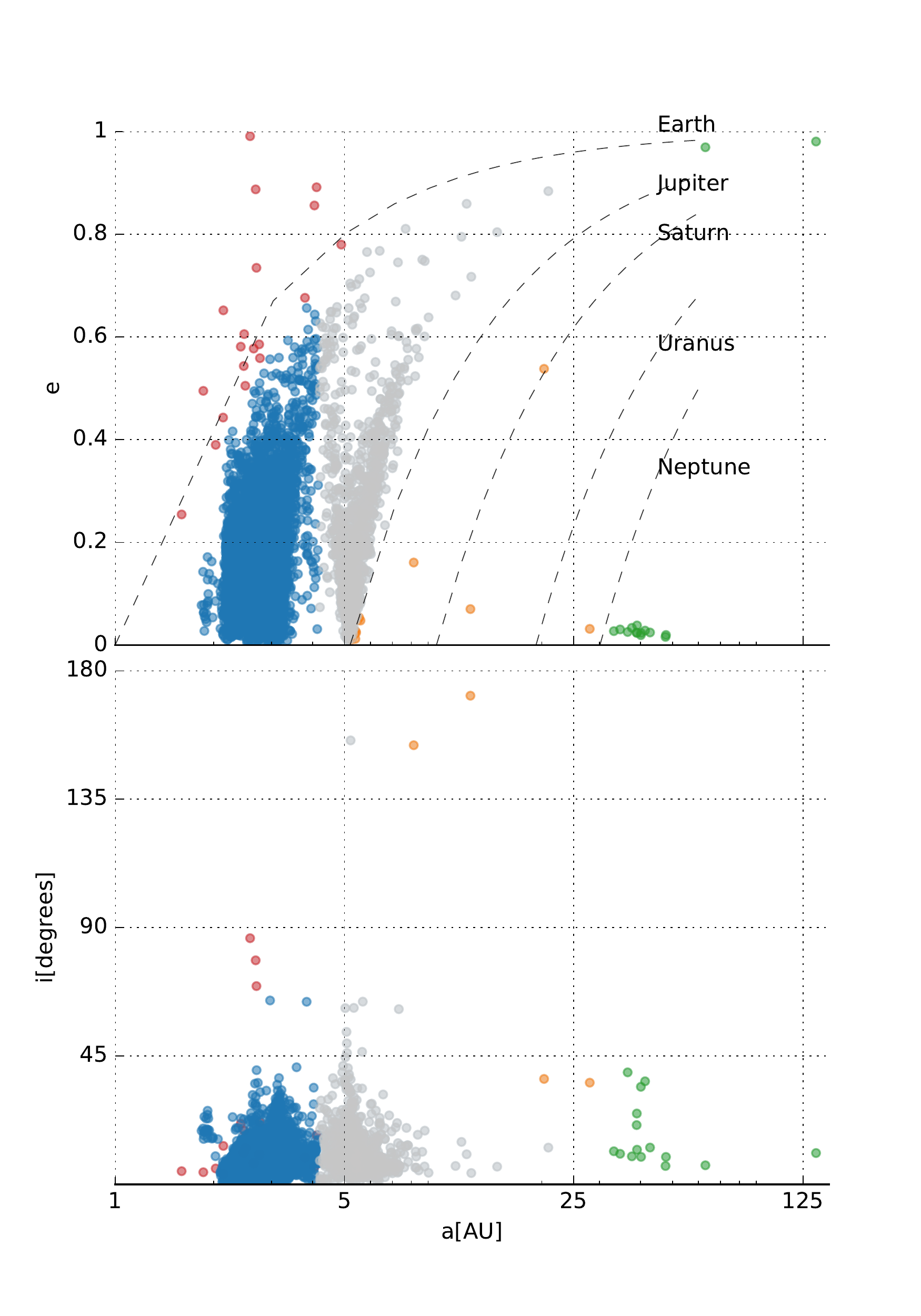}
   \caption{Orbital solution for all tracks that yield bound orbits and a maximum deviation of 2 arc-seconds from the model. There are 7,700 objects in 2014 that fulfill this criterion. The lines show the solutions that share their pericenter distance with the outer planets. We show Near Earth Objects in red, Main Belt asteroids in blue, Centaurs in orange, trans-Neptunian objects in green and those that did not fit any of the other criteria in gray, most likely Jupiter Trojans for which the uncertainties on their orbits do not allow us to tag them as such.}
   \label{fig:orbits}
    \end{figure}

This resulted in 7,700 bound trajectories\footnote{Available in \url{http://www.das.uchile.cl/~jpena/HiTS_2014/}} with a typical arc of 2.2 days, which are shown in Figure \ref{fig:orbits}. Most of the asteroids found (6,687) belong to the Main Belt (in blue, perihelion $q>1.3$ AU and semi-major axis $a<4.2$ AU); there are 19 Near-Earth objects (NEOs, red, $q < 1.3$ AU); 14 Centaurs (in orange, $q> 5.2$ AU and $a<30$ AU); and 15 trans-neptunian objects (TNOs, green, $a>30$ AU). We use orbital parameter limits from the literature \citep{Gladman.2008,Parker.2008,Juric.2002}. Note that there are 708 unclassified objects (in gray) which are outside the regions of these other families. These are most likely Jupiter Trojans for which a $\sim$2--day arc is insufficient to constrain their orbital parameters precisely enough to be classified as such.

\begin{figure}
   \centering
   \includegraphics[width=\hsize]{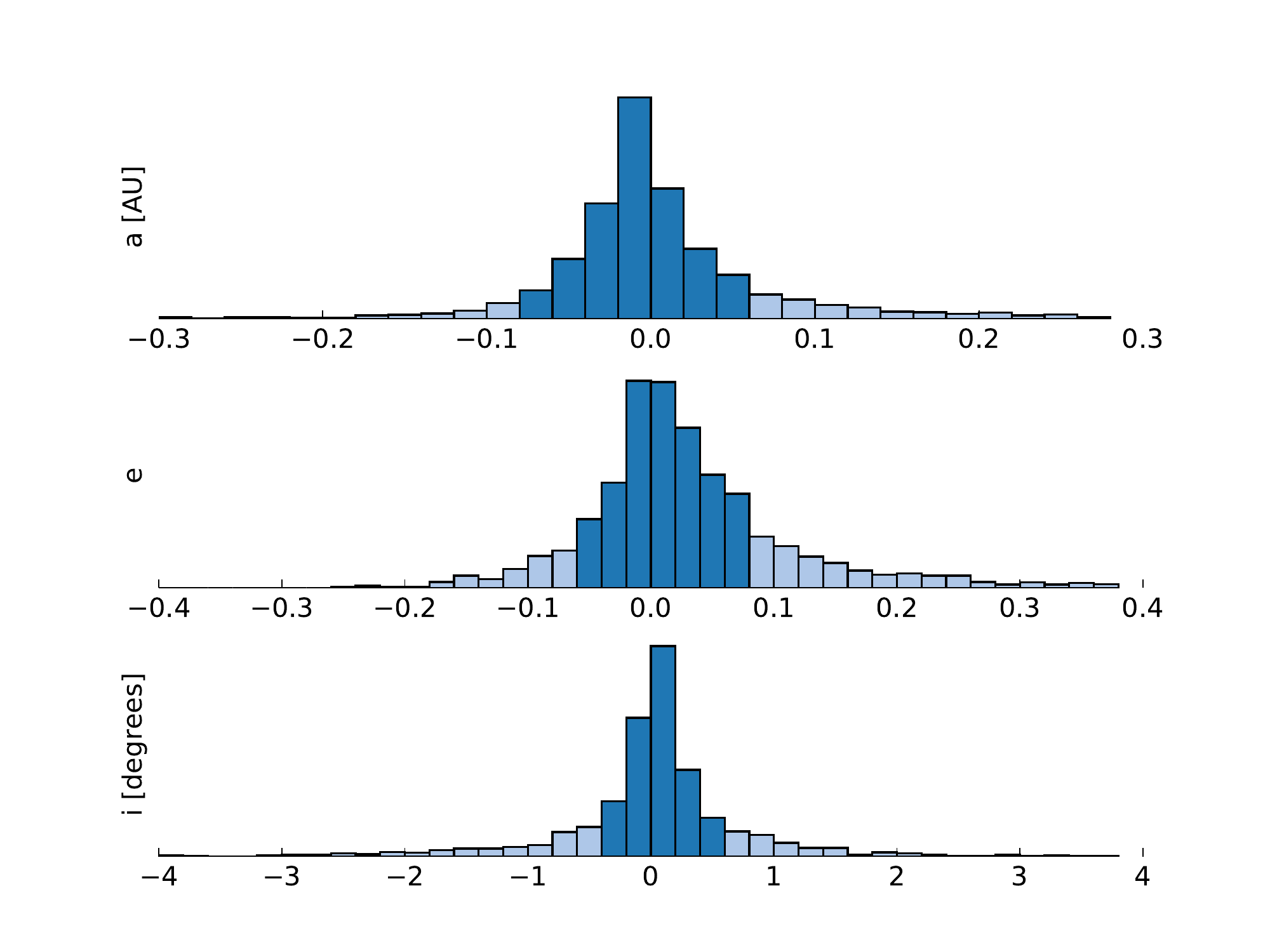}
   \caption{Errors for the estimated orbits of known asteroids in our sample. We highlight the $1-\sigma$ confidence region. From top to bottom we show errors in semi-major axis, eccentricity, and inclination. The implied $1-\sigma$ confidence region for our orbital solutions is: $\sigma_a = [-0.08, 0.06]~{\rm AU}$, $\sigma_e = [-0.06, 0.08]$, and $\sigma_i = [-0.4,0.6]~{\rm degrees}$.
}
\label{fig:orberrors}%
\end{figure}

We compute the uncertainty in the fitted orbital parameters as the standard deviation of the errors between the orbital parameters derived from our fit and those reported by the MPC (see Figure~\ref{fig:orberrors}). In this analysis we did not consider objects observed only in a single night as their arc does not allow a reliable fit, leaving orbital parameters of 2,464 known asteroids to compare with. We report our $1-\sigma$ uncertainties as the interval that bounds 68\% of the errors around the mode (as in the normal distribution) to be $\sigma_a\sim 0.07 ~{\rm AU}$ for the semi-major axis, $\sigma_{\rm e} \sim 0.07 $ for the eccentricity and $\sigma_i\sim 0.5~ {\rm deg}$ for the inclination.

Most of the computed orbital parameters show errors that follow a normal distribution around the known value (See Figure \ref{fig:orberrors}). However, there is a tail of fitted orbits that fall far outside the range of that figure. These orbits correspond mostly to those consistent with Trojan orbits (in gray in Figure \ref{fig:orbits}). The fraction of those poor orbit fits are 7\%, 1\% and 1\% for $a$, $e$ and $i$ respectively.

\subsection{Color}

\begin{figure}
   \centering
   \includegraphics[width=\hsize]{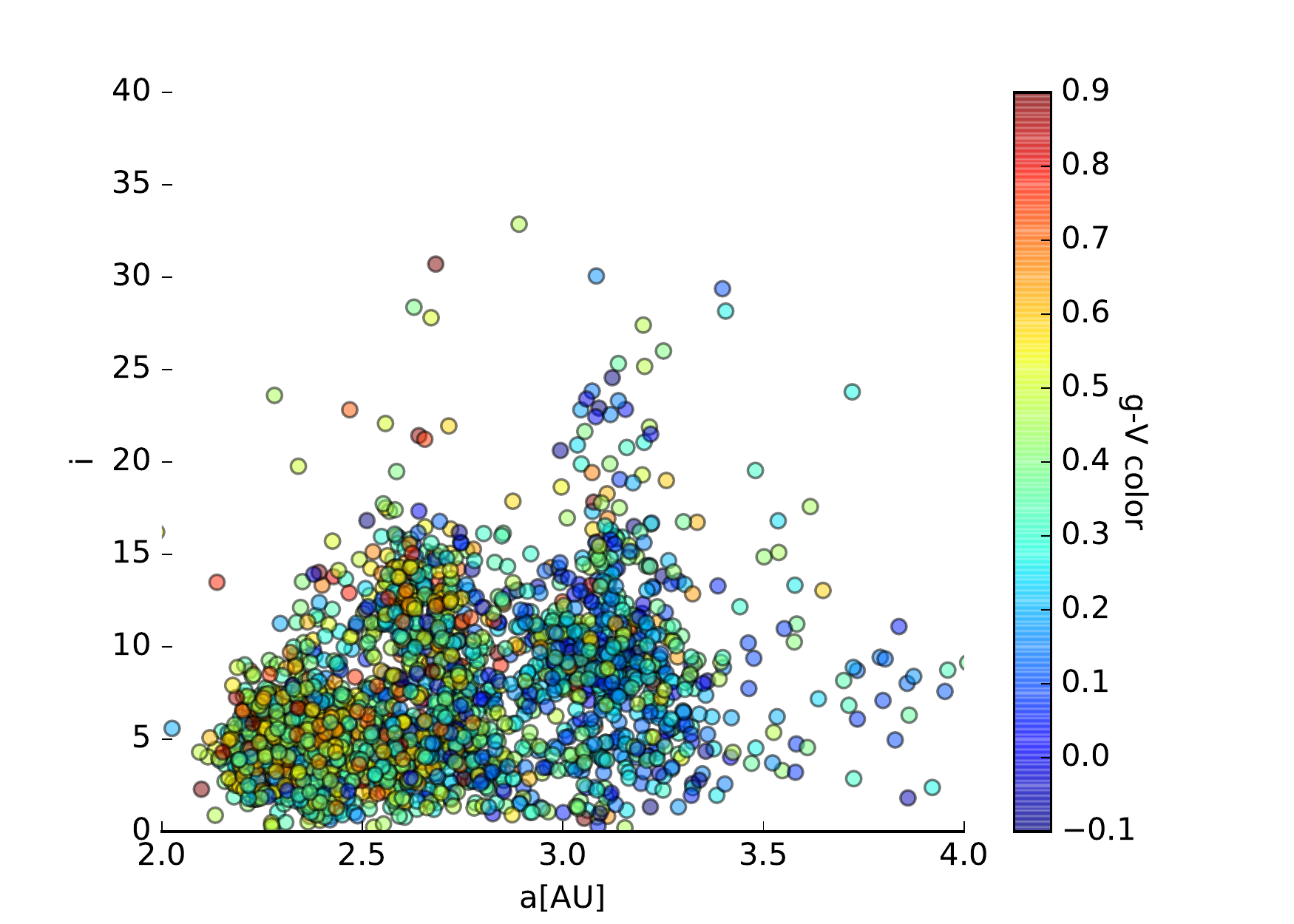}
   \caption{Known Main Belt asteroids rediscovered in this survey with arcs longer than one day are shown in this $g-V$ color vs orbital parameters $a$ and $i$, as inferred from our photometry and orbital fit.}
\label{fig:colors}%
\end{figure}

Using our own $g$ magnitudes with the V magnitudes provided by MPC, we calculate the colors of those objects considered in Figure \ref{fig:orberrors} that could be fit into a Keplerian orbit and consistent with Main Belt asteroids, as shown in Figure \ref{fig:colors}. Asteroids at the Inner Belt are redder than those in the Outer Belt, as seen in Figure 4 of \cite{Parker.2008}. 
This exercise shows how adding observations with another filter would allow us to extend the color-family relationships in the Main Belt \citep{Parker.2008} to much smaller objects.

\section{Discussion and Conclusions}

%
\begin{table*}
\centering                          
\caption{Surveys with Minor Planets Detections}           
\begin{tabular}{l c c c c c}        
\hline\hline                 
Survey & $\Omega$ & lim. mag.\tablenotemark{*} & $N_\mathrm{obj}$ & Strategy\\    
& deg$^2$ & & & \\    
\hline                        

SDSS - I/II\tablenotemark{a} & $\sim$10,000 & $r^{*}\sim 22$ & 471,569 asteroids\tablenotemark{b} & $\sim 8$ years survey.\\
 & & & & Each field in 5 bands (72s between bands). \\
 & & & & 28$\%$ of surveyed area is covered twice or more.\tablenotemark{c} \\
 & & & & \\
 
CFEPS\tablenotemark{d} & $\sim$320 & $m_g \sim 24$ & 169 TNOs & $\sim$4 years for discovery + follow-up 2 years later. \\
 & & & & Cadence optimized for TNOs (few observations \\
 & & & & spanned on days to months). \\
 & & & & \\

WISE\tablenotemark{e} & All the sky & $W1 \sim$15.3\tablenotemark{f} & $\sim$150,000\tablenotemark{g} & $\sim$1.5 years survey. All the sky is observed\\
 & & & & in $\sim$7 months, visiting the same area at least\\
 & & & & 8 times (on $\sim$10 days).\\
 & & & & \\
 
(i)PTF\tablenotemark{h} & $\sim$10,000 & R$\sim$20.5 & $\sim$1582\tablenotemark{i} & $\sim$8 years survey. Cadences vary from minutes\\
 & & & (new bodies) & to days. Known asteroids are extracted\\
 & & & & before linking. Discoveries are made via\\
 & & & & linking and streak-detections (for NEOs)\\
 & & & & \\

PS-1\tablenotemark{j} & $\sim$30,000 & $r_{P1}\sim 22$ & 600,000 asteroids\tablenotemark{g} & $\sim 4$ years survey. Different observing methods. \\
 & & $w_{P1}\sim 22.5$ & & Most of the surveyed area is observed 4 times\\ 
 & & & & a year. There are also observations at the\\
 & & & &  ecliptic plane for NEOs and KBOs.\\
 & & & & \\

HiTS\tablenotemark{k} & 120 & $g_{50}$=23.5-24.5 &  7,700 & $\sim$5 days survey. Each surveyed area \\
 & & & & is visited 4-5 times per night every 2 hours.\\
 & & & & \\

LSST\tablenotemark{l} & $\sim$25,000 & $r\sim$24.5 & $>$ 5 millions & $\sim$10 years survey. 2 pairs of visits per field \\
 & & & (estimated) & separated by $\sim$30 mins. covering the entire \\
 & & & & visible sky every 3-4 days ($\sim$ 4,000 deg$^2$\tablenotemark{m}).\\
 & & & & \\
\hline                                   
\end{tabular}
\tablenotetext{*}{\small{Most surveys quote the limiting magnitude (LM) for individual detections. The LM for asteroids requires a control sample. In our survey we estimate our LM with known asteroids ($g_{50}\sim23.5$) as shown in Figure \ref{fig:maghist} and \ref{fig:eff}}}
\tablenotetext{a}{\small{\cite{Ivezic.2001}}}
\tablenotetext{b}{\small{\url{http://faculty.washington.edu/ivezic/sdssmoc/sdssmoc.html} ; \cite{Ivezic.2010}}}
\tablenotetext{c}{\small{The Sloan Digital Sky Survey Project Book (\url{http://www.astro.princeton.edu/PBOOK/})}}
\tablenotetext{d}{\small{\cite{Petit.2011, Kavelaars.2009}}}
\tablenotetext{e}{\small{\cite{Mainzer.2011, Wright.2010}}}
\tablenotetext{f}{\small{\url{http://wise2.ipac.caltech.edu/docs/release/prelim/expsup/sec2_2.html}}}
\tablenotetext{g}{\small{\cite{Veres.2017}}}
\tablenotetext{h}{\small{\cite{Law.2009, Kulkarni.2013, Waszczak.2013, Waszczak.2015, Waszczak.2017}}}
\tablenotetext{i}{\small{\url{https://www.ptf.caltech.edu/page/asteroids_data}}}
\tablenotetext{j}{\small{\cite{Chambers.2016, Lin.2016}}}
\tablenotetext{k}{\small{\cite{2016ApJ...832..155F}}}
\tablenotetext{l}{\small{\cite{Jones.2016}}}
\tablenotetext{m}{\small{Assuming 2 visits of 30 seconds per 10 hr night}}

\label{table:survey}      
\end{table*}

We report on our search for asteroids and other Solar System populations imaged in the High cadence Transient Survey (HiTS). This wide-field survey covered $\sim 120$ square degrees over 6 nights in 2014 looking for time variable phenomena, in particular Supernovae. We took advantage of the readily available HiTS's pipeline's variable source catalog (F16), already classified by machine learning, and performed two different search algorithms for motion consistent with known small-body populations.
We found 7,700 viable orbits, identifying 19 NEOs, 6,687 Main Belt asteroids (around 2,500 previously known), 14 Centaurs, and 15 TNOs  as shown in Figure \ref{fig:orbits}. 
It is important to notice that this characterization is based on relatively short arcs (mean arc $\sim 2.2$ days). Our orbits are precise enough to determine orbital parameters with precision for the longest arcs, but not enough to classify all reported objects (see Figure \ref{fig:orberrors}). We used the list of known objects from the MPC to check on our efficiency and orbital accuracy. Of those known orbits in our tracks (3,812), only those with arcs longer than a night (2,464) yielded bound orbits. There were 708 objects (gray dots in Figure \ref{fig:orbits}) with orbits mostly consistent with  Jupiter Trojans but with large orbital uncertainties. 

Our results serve as a good model for the asteroid discovery efficiency that can be expected from synoptic surveys like LSST, with one cadence optimized for several different science drivers \citep{LSST.2009, Jones.2016}. Table \ref{table:survey} shows a comparison of this work with other past and future Solar System searches on wide-field surveys. We include the total surveyed area, limiting magnitude, number of discoveries, and timespan for data gathering as a basis of comparison. Cadences vary greatly between surveys, and usually within the same survey. LSST's basic survey strategy of observing one field twice a night, covering the entire sky every 3-4 days is similar to HiTS' constant cadence returning to each field over 5 nights. This match between observing cadences yields a similar linking problem, that requires finding
tracklets within a few days to form a proper Solar System trajectory.

Our results are enabled by the overwhelming rejection of spurious detections by the machine learning algorithm described in \cite{Cabrera.2017} and F16. This filtering process on raw detections not only reduces the execution time, but also reduces false discoveries and the number of trajectories contaminated by incorrect detections (which scales with the number density of detections). This ``track confusion'' increases the errors in orbital parameters and is an important problem when trying to extend one track into another night. LSST is expected to produce $\sim$10 million alerts per night \citep{LSST.2009}. It would take $\sim$6 hours in a single 2.2 GHz processor to compute Machine Learning features for all of these. This problem is readily parallelizable to substantially reduce the extra processing time.

Although this classification algorithm (F16) is not tailored for tracked moving objects, only 5$\%$ of detections and 2$\%$ of real traceable objects are rejected. Future work is needed in describing sources that deviate from a point source to better account for this obvious source of confusion, as well as other special cases like binaries, comets, etc.
Having a well-defined probability for each detection in the source list of any future survey will improve the accuracy of any linking algorithm, especially when dealing with track confusion.
We can use the surveyed area to scale the computational cost of our search if applied to LSST assuming they use a vetting algorithm that rejects $80\%$ of the alerts. Our simple ${\cal O}(n^{2})$ search algorithm with no major optimization would be able to link the area surveyed in one night to data obtained 3-4 nights before in 36~hrs (using 120 nodes of 20 2.2~GHz cores each). However the ${\cal O}(n\log{n})$ algorithm would only take 2 hours in a single node of 20 cores.
This vetting of detections would enable users to run their own linking algorithms to search for moving targets in the LSST's alert stream (the analogue of our transient candidates catalogue).
We emphasize that improving and optimizing the classification of raw non-vetted detections should yield great benefits for the implementation of more general search algorithms.
We have shown that using a similar ML classification scheme, LSST could provide better alerts with little extra overhead and that simple algorithms can then link moving objects with an arc of 3-4 days within hours. This arc is long enough to constrain the orbital parameters to a few percent for asteroids observed near opposition. Adding ML to the reduction process means that a fast-moving object could be discovered and followed up on the same night.

As wide-field surveys get deeper, understanding their linking efficiencies and biases will allow the size distribution of asteroids to extend to smaller sizes (Figure \ref{fig:eff}). The HiTS 2015 campaign provides photometry mainly in $g$ band but also in the $r$ band. A similar treatment of that data will yield colors for fainter and smaller asteroids. Studying the color distribution as a function of size can be linked to the dynamical history of these populations.

\section{Acknowledgements}
J.P., C.F., F.F., J.S.M., G.C.V., S.G.G. and J.M. acknowledge support from Basal Project PFB-03, Centro de Modelamiento Matem\'atico (CMM), Universidad de Chile..
F.F., M.H., S.G.G, P.A.E and J.M. acknowledge support from the Ministry of Economy, Development, and Tourism’s Millennium Science Initiative through grant IC120009, awarded to The Millennium Institute of Astrophysics (MAS).
C.F. acknowledges support from the BASAL Centro de Astrof\'isica y Tecnolog\'ias Afines (CATA) PFB-06/2007.
F.F. acknowledges support from Conicyt through the Fondecyt Initiation into Research project No. 11130228.
S.G.G., P.H., G.C.V and L.G. acknowledge support from FONDECYT postdoctoral grants 3130680, 3150460, 3160747 and 3140566, respectively.
F.F., J.C.M., P.H., G.C.V., and P.A.E. acknowledge support from Conicyt through the Programme of International Cooperation project. 
L.G. was supported in part by the US National Science Foundation under Grant AST-1311862.
F.F., J.C.M., P.H., G.C.V, P.A.E. acknowledge support from Conicyt through the Programme of International Cooperation project DPI20140090.
J.M. acknowledges the support from CONICYT Chile through (CONICYT-PCHA/Doctorado-Nacional/2014-21140892).
Powered{@}NLHPC: this research was partially supported by the supercomputing infrastructure of the NLHPC (ECM-02).
Part of this work was done under the Harvard-Chile data science school.
This project used data obtained with the Dark Energy Camera (DECam), which was constructed by the Dark Energy Survey (DES) collaboration.
Funding for the DES Projects has been provided by the U.S. Department of Energy, the U.S. National Science Foundation, the Ministry of Science and Education of Spain, the Science and Technology Facilities Council of the United Kingdom, the Higher Education Funding Council for England, the National Center for Supercomputing Applications at the University of Illinois at Urbana--Champaign, the Kavli Institute of Cosmological Physics at the University of Chicago, Center for Cosmology and Astro--Particle Physics at the Ohio State University, the Mitchell Institute for Fundamental Physics and Astronomy at Texas A\&M University, Financiadora de Estudos e Projetos, Funda\c{c}\~ao Carlos Chagas Filho de Amparo, Financiadora de Estudos e Projetos, Funda\c{c}\~ao Carlos Chagas Filho de Amparo \`a Pesquisa do Estado do Rio de Janeiro, Conselho Nacional de Desenvolvimento Cient\'ifico e Tecnol\'ogico and the Minist\'erio da Ci\^encia, Tecnologia e Inova\c{c}\~ao, the Deutsche Forschungsgemeinschaft and the Collaborating Institutions in the Dark Energy Survey. 
The Collaborating Institutions are Argonne National Laboratory, the University of California at Santa Cruz, the University of Cambridge, Centro de Investigaciones Energ\'eticas, Medioambientales y Tecnol\'ogicas--Madrid, the University of Chicago, University College London, the DES--Brazil Consortium, the University of Edinburgh, the Eidgen\"ossische Technische Hochschule (ETH) Z\"urich, Fermi National Accelerator Laboratory, the University of Illinois at Urbana--Champaign, the Institut de Ci\`encies de l'Espai (IEEC/CSIC), the Institut de F\'isica d'Altes Energies, Lawrence Berkeley National Laboratory, the Ludwig--Maximilians Universit\"at M\"unchen and the associated Excellence Cluster Universe, the University of Michigan, the National Optical Astronomy Observatory, the University of Nottingham, the Ohio State University, the University of Pennsylvania, the University of Portsmouth, SLAC National Accelerator Laboratory, Stanford University, the University of Sussex, and Texas A\&M. University. 
J.P. and F.F. acknowledge Pavlos Protopapas for useful discussion. J.P. acknowledge Michael Lackner for useful discussion.

\end{document}